\documentclass[preprint,journal]{vgtc}            

\onlineid{1629}

\vgtccategory{Research}

\vgtcpapertype{application}

\author{%
    \authororcid{Wei Zhang}{0000-0002-8321-4607},
    \authororcid{Wong Kam-Kwai}{0000-0002-2813-1972}, 
    Biying Xu,
    Yiwen Ren,
    Yuhuai Li, 
    Minfeng Zhu, 
    Yingchaojie Feng, 
    and Wei Chen
}

\authorfooter{
  \item W. Zhang, B. Yu, Y. Ren, Y. Li, M. Zhu, Y. Feng, and W. Chen are with the State Key Lab of CAD\&CG, Zhejiang University. Wei Chen (corresponding author) is also with the Laboratory of Art and Archaeology Image (Zhejiang University), Ministry of Education, China. E-mail: \{zw\_yixian, biyingx, renyiwen, yuhuaili, minfeng\_zhu, fycj, chenvis\}@zju.edu.cn.
  \item W. Kam-Kwai is with Hong Kong University of Science and Technology and Georgia Institute of Technology. Email: kkwongar@connect.ust.hk.
}

\UseRawInputEncoding

\abstract{
The integration of new technology with cultural studies enhances our understanding of cultural heritage but often struggles to connect with diverse audiences. 
It is challenging to align personal interpretations with the intended meanings across different cultures. 
Our study investigates the important factors in appreciating art from a cross-cultural perspective.
We explore the application of Large Language Models (LLMs) to bridge the cultural and language barriers in understanding Traditional Chinese Paintings (TCPs). 
We present CultiVerse, a visual analytics system that utilizes LLMs within a mixed-initiative framework, enhancing interpretative appreciation of TCP in a cross-cultural dialogue.
CultiVerse addresses the challenge of translating the nuanced symbolism in art, which involves interpreting complex cultural contexts, aligning cross-cultural symbols, and validating cultural acceptance.
CultiVerse integrates an interactive interface with the analytical capability of LLMs to explore a curated TCP dataset, facilitating the analysis of multifaceted symbolic meanings and the exploration of cross-cultural serendipitous discoveries.
Empirical evaluations affirm that CultiVerse significantly improves cross-cultural understanding, offering deeper insights and engaging art appreciation.}

\title{CultiVerse: Towards Cross-Cultural Understanding for Paintings \texorpdfstring{\\}{} with Large Language Model}


\teaser{
  \centering
 \includegraphics[width=\linewidth]{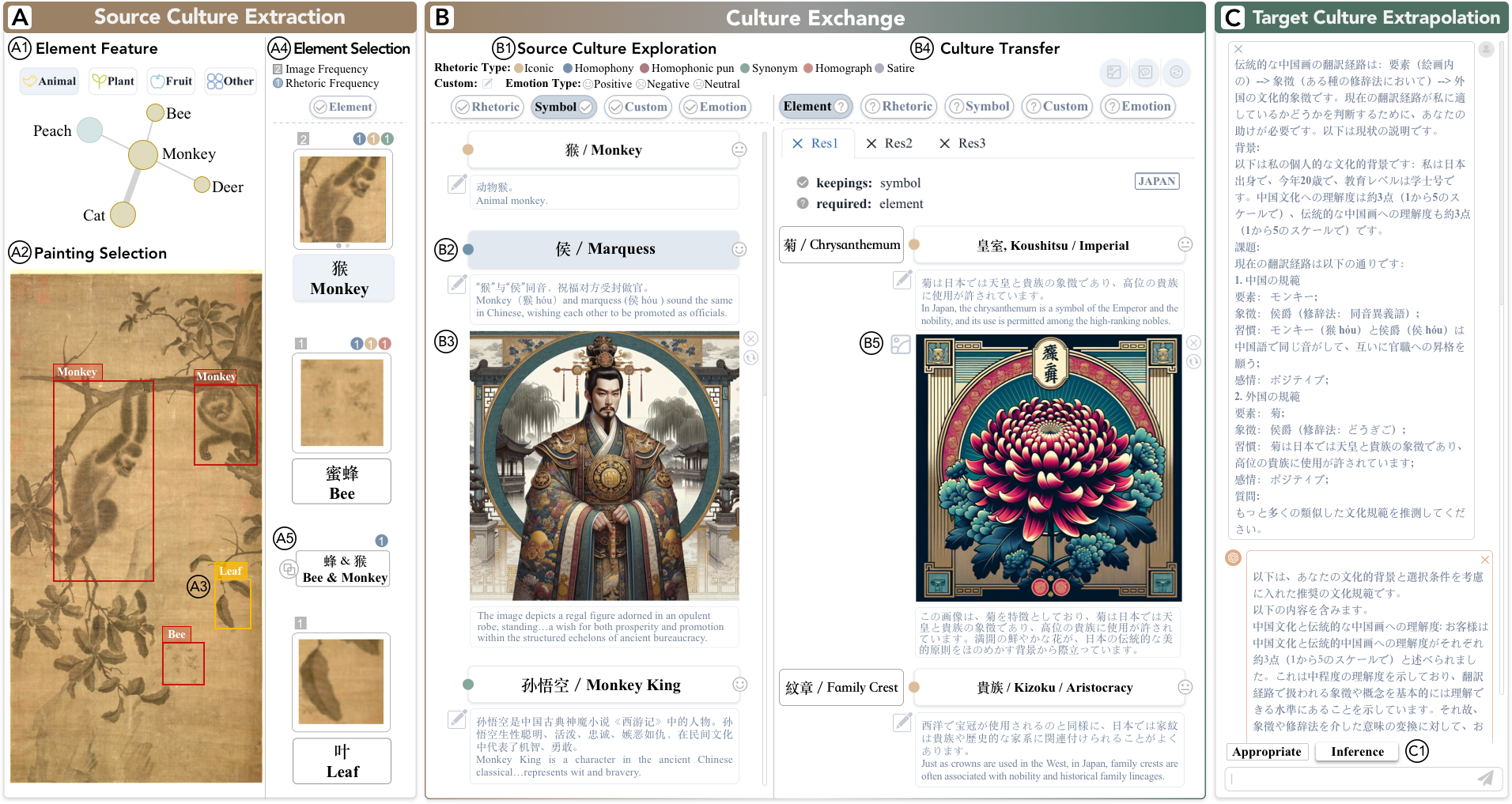}
  \caption{The interface of CultiVerse offers a streamlined approach to exploring and appreciating paints for enhanced cross-cultural understanding. It starts with the source culture extraction (A), which provides accurate information about the TCP for identifying and associating visual elements. Next, the culture exchange (B) shows elements and symbols across cultures, aiming to align the disparities in diverse cultural norms and provide LLM-generated content to support multimodal understanding. Lastly, the target culture extrapolation (C) encourages curiosity-led exploration while ensuring adherence to cultural norms and appropriateness.}
  \label{fig:teaser}
}

\graphicspath{{figs/}{figures/}{pictures/}{images/}{./}} 

\usepackage{tabu}                      
\usepackage{booktabs}                  
\usepackage{lipsum}                    
\usepackage{mwe}                       

\usepackage{mathptmx}                  

\usepackage{xspace,xpunctuate}

\newcommand{\ie}{\textit{i.e.},\xspace}
\newcommand{\etal}{\xspace\textit{et al.}\xspace}
\newcommand{\eg}{\textit{e.g.},\xspace}

\newcommand{\rc}[1]{\textcolor{black}{#1}}
\newcommand{\rw}[1]{\textcolor{black}{#1}}

\usepackage[svgnames]{xcolor}
\usepackage{wrapfig}

\usepackage{CJKutf8} 
\usepackage{tipa}

\definecolor{mybackground}{RGB}{237,240,239} 
\definecolor{myline}{RGB}{57,95,80} 

\definecolor{mybackgroundA}{RGB}{243,241,239} 
\definecolor{mylineA}{RGB}{155,130,103} 

\definecolor{topcolor}{RGB}{243,241,239}
\definecolor{midcolor}{RGB}{255,255,255}
\definecolor{bottomcolor}{RGB}{237,240,239}
\definecolor{linetop}{RGB}{155,130,103}
\definecolor{linemid}{RGB}{186,186,186}
\definecolor{linebottom}{RGB}{78,112,99}

\usepackage{tikz}  
\definecolor{myfill}{rgb}{0.31, 0.44, 0.39}  
\definecolor{mytext}{rgb}{1, 1, 1}

\definecolor{jingtailan}{RGB}{39,117,182}

\usepackage[utf8]{inputenc}
\usepackage{graphicx}
\usepackage{wrapfig}
\usepackage[most]{tcolorbox}
\usepackage{float}  
\usepackage{tikz}  
\usetikzlibrary{shadings}
\usepackage{pifont} 

\definecolor{Iconic}{HTML}{D6BF9E}
\definecolor{Homophony}{HTML}{7891AA}
\definecolor{Homophonic pun}{HTML}{AD7982}
\definecolor{Synonym}{HTML}{8AA79B}
\definecolor{Homograph}{HTML}{CA9087}
\definecolor{Satire}{HTML}{AAB0BE}

\definecolor{Plant}{HTML}{C0CC32}
\definecolor{Animal}{HTML}{E8AF05}
\definecolor{Fruit}{HTML}{71B8B8}
\definecolor{Others}{HTML}{6084BE}

\definecolor{Political}{HTML}{7891AA}
\definecolor{Academic}{HTML}{AD7982}
\definecolor{Social}{HTML}{8AA79B}
\definecolor{Kinship}{HTML}{CA9087}
\definecolor{Paint}{HTML}{D6BF9E}
\definecolor{Painter}{HTML}{A9906D}

\definecolor{linkeddata}{HTML}{000000}
\definecolor{biography}{HTML}{000000}
\definecolor{handscrollfeature}{HTML}{000000}
\definecolor{hovera}{HTML}{000000}
\definecolor{hoverb}{HTML}{000000}
\definecolor{hoverc}{HTML}{000000}
\definecolor{hoverd}{HTML}{000000}

\newcommand{\eventtype}[2]{\includegraphics[height=\fontcharht\font`\B]{figs/eventtype/#1.pdf}\textit{\textcolor{#1}{#2}}}

\usepackage{amsmath}
\usepackage{bm}

\mathchardef\dash="2D

\begin{document}



\maketitle

\section{Introduction}
The intersection of technology and cultural studies has significantly reshaped the analysis of historical narratives through modern perspectives~\cite{zhang2023uncertainty,zhang2023cohortva,Zhang2024Scroll}.
These advanced studies, rich in historical insight, facilitate in-depth exploration of cultural heritages~\cite{zhang2021visual,feng2022ipoet,wang2021interactive}.
Yet, the intricate design of such tools is often tailored for experts, limiting their accessibility to reach a broader audience who may lack essential prior knowledge.
The issue is exacerbated for individuals from different cultural backgrounds.
They may not be aware of the necessary historical and cultural contexts to fully appreciate the subtle meanings embedded in cultural artifacts. 
Without adequate guidance, there is a risk that the intended implications could be misinterpreted or distorted with biased perceptions.
This scenario motivates the pressing need to create systems that make insightful cultural analyses accessible to a broader audience regardless of their prior knowledge or cultural backgrounds.

In this context, cross-cultural understanding is crucial in fostering the ability to recognize, interpret, and appropriately respond to cultural phenomena (\eg language, customs, social norms, and art) that seem unfamiliar or exotic due to cultural differences. 
This understanding is essential for nurturing effective intercultural communication and enhancing multicultural awareness and sensitivity. 
Our study leverages Traditional Chinese Paintings (TCP) as a case in point to deepen cross-cultural understanding. 
The visual aspect of these paintings transcends spoken and written language barriers, offering a universally accessible medium to explore and understand the underlying values of another culture. 
By engaging with and interpreting artworks from varied cultural backgrounds, individuals can appreciate different cultures' distinctive perspectives and ways of life, laying a foundation for mutual understanding. 
Fostering cross-cultural understanding not only enriches individual perspectives but also contributes to the promotion of broader cultural exchange and harmony.

We explore using Large Language Models (LLMs) to bridge cultural and language barriers for cross-cultural understanding. 
Pretrained on a diverse corpus of high-quality texts from different cultures, these models are adept at analyzing and contextualizing complex information. 
They also synergize with generative image models, enabling image creation in diverse artistic styles upon request, which aids in visualizing complex concepts that are difficult to articulate with words alone.
Utilizing LLMs offers unprecedented opportunities to enhance the cross-cultural understanding of paintings, making painting appreciation more inclusive to people from different cultural backgrounds.

\begin{figure}[t]
\centering
\includegraphics[width=\linewidth]{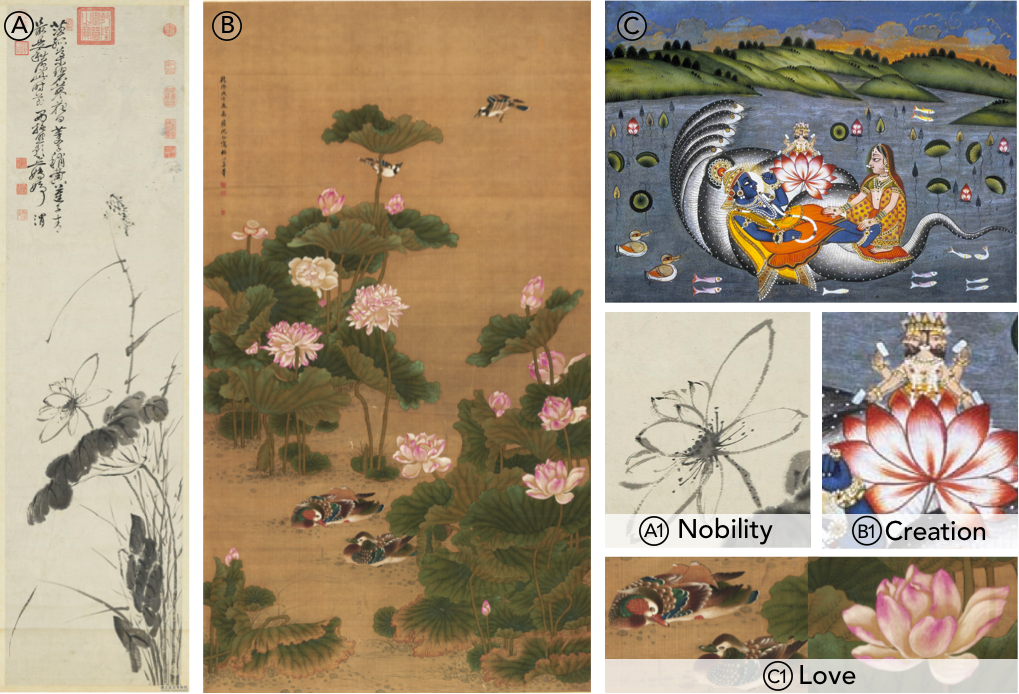}
\caption{The lotus flower carries diverse meanings in different cultural contexts. (A) The``\textit{Lotus Painting}'' by Xu Wei symbolizes nobility~\cite{ruan2023lotus}. (B) The ``\textit{Lotus Pond and Mandarin Ducks}'' by Shen Quan expresses a harmonious love relationship~\cite{sun2023lotus}. (C) Painting from the Bhagavata Purana, a Hindu manuscript, describes creation~\cite{hallman1954lotus}.}
\label{fig:abstract}
\vspace{-6pt}
\end{figure}

However, in developing system prototypes, we have identified three key challenges in harnessing LLMs for such purposes. 
We illustrate them using the lotus flower as an example. 
In TCP, it symbolizes purity and spiritual enlightenment when portrayed alone~\cite{ruan2023lotus} (\cref{fig:abstract}A), while its meaning shifts to represent a harmonious love relationship when associated with mandarin ducks~\cite{sun2023lotus} (\cref{fig:abstract}B). 
This shift exemplifies the challenge of contextual interpretation within the multifaceted symbolic meanings. 
In contrast, the lotus suggests a different connotation of divinity and creation in Hindu art~\cite{hallman1954lotus} (\cref{fig:abstract}C), and signifies rebirth and restoration in Ancient Egyptian culture~\cite{taylor2010lotus}. 
This highlights the challenge of cross-cultural symbol alignment. 
Furthermore, the dragon provides another illustrative example that is respected in Chinese culture as a symbol of strength and prosperity, but is often associated with threat and malevolence in Christian culture~\cite{yuan2015cultural}. 
This divergence highlights the difficulty in validating variations in different cultural customs. 
Despite their proficiency in processing textual information, LLMs exhibit limitations in accurately translating the rich visual narratives of cultural arts.
Human intervention is required to guide the sense-making process and prevent cultural misinterpretations or oversimplifications.

We introduce CultiVerse~\cite{CultiVerse}, a visual analytics system that harnesses LLMs in a mixed-initiative framework, designed for novices to enhance the interpretive engagement with cultural artifacts in a cross-cultural dialogue. 
We first characterize the domain concept of ``cultural norm'' and curate a corresponding TCP dataset. 
This semi-automatically annotated dataset captures the broad spectrum of symbolic meanings associated with various elements depicted in TCP. 
Then, we design an interactive interface that visualizes the multifaceted symbolic meanings and intercultural connections, enabling users to filter and analyze the cultural norms effectively. 
The interface seamlessly integrates LLMs to augment the cross-cultural dialogue with enriched contextual understanding, align cross-cultural interpretations, and propose alternative associative ideas to inspire serendipitous discoveries. 
Furthermore, we have incorporated pre-configured structured prompts to assist users in efficiently validating the cultural acceptability of their exploratory findings. 
CultiVerse's mixed-initiative workflow progressively guides both the user and the LLM towards generating more refined prompts and results, thus optimizing the interpretive accuracy. 
Through comparative studies and user interviews, we have demonstrated that CultiVerse significantly improves the efficiency of cross-cultural understanding and yields more varied insights than conventional methods. 
Our system not only streamlines the analytical process but also deepens the level of interpretative exploration, providing transformative perspectives into the interplay of culture and art.
This study has the following contributions:
\begin{itemize}[noitemsep,topsep=1pt]
    \item{We characterize the problem of cross-cultural understanding for TCP and curate a TCP Cultural Norm Dataset that captures the diverse symbolic meanings of elements.}
    \item{We develop CultiVerse based on the proposed mixed-initiative workflow, leveraging visual analytic approaches and LLMs.}
    \item{We conduct comparative user studies and interviews to demonstrate the effectiveness and usefulness of CultiVerse.}
\end{itemize}

\section{Related works}

\subsection{Visual analytics for cultural understanding}

Cultural understanding is a complex process that involves recognizing, appreciating, and interpreting specific cultural phenomena~\cite{rohmah2021cross}.
Cultural artifacts such as ancient books~\cite{zhang2023cohortva}, ancient pottery~\cite{Ye2023Puzzle}, and paintings~\cite{chen2023painting} play a crucial role in cultural understanding. 
Researchers use digital tools to uncover the cultural phenomena hidden behind these artifacts. The rapid development of digital humanities in recent years demonstrates that visual analytics is an effective method to enhance cultural understanding and improve its efficiency~\cite{8617736}. 
For example, Zhang\etal~\cite{zhang2021visual} used visual analytics to delve deeper into the social-cultural context of the Song Dynasty through poetry.
LiberRoad~\cite{2024LiberRoad} used ancient books as a medium to visualize the international dissemination of Chinese classical literature, promoting cultural exchange and integration between China and Japan.
ScrollTimes~\cite{Zhang2024Scroll} facilitated art historians in tracing the provenance of paintings by extracting elements from these paintings and conducting visual associative analyses.

However, these existing works often target a niche audience of experts for in-depth historical analysis, requiring extensive domain knowledge and often overlooking the broader goal of making cultural art and phenomena more accessible to the general public. To bridge this gap, we have designed and developed a visual analytic system, CultiVerse, to bring insights related to cultural symbols to a general audience, particularly those from diverse cultural backgrounds and those interested in learning about other cultures.

\subsection{Facilitating cross-cultural understanding}
Cross-cultural understanding extends the definition of cultural understanding to include language, customs, social norms, and artistic expressions that might be unfamiliar due to cultural discrepancies~\cite{rohmah2021cross,Letts2015Cross, Roth2001Material}. 
This understanding is crucial for effective communication and interaction in our increasingly globalized world~\cite{2021IMPROVING}. 

Amanda~\cite{Harris2014Archival} pointed out that in the absence of a common language, painting can serve as vital modes of communication for understanding other cultures.
Liu~\cite{Liu2009Cross} discussed the methodology used in the author's paintings to visualize cross-culturalism.
Emilie~\cite{Martinez2013Listen} investigated a cross-cultural experiment that combined classical music and soundscapes with painting to enhance intercultural competence.
Zhao\etal~\cite{mti2020016} designed a tablet-based authoring tool to support cross-cultural appreciation of TCP through interactive experiences. Similarly, Pan\etal~\cite{pan2020discussion} examined the aesthetic principles in TCP and Western cultural active factors, aiming to deconstruct and distill cultural elements for modern product design processes. 
However, these works mainly focus on appreciating the artistic style of paintings, lacking in-depth translation and interpretation of the cultural meanings conveyed by the paintings.
Our work, through meticulous requirement research, outlines a detailed analytical process for cross-cultural understanding based on paintings for audiences from diverse cultural backgrounds.

Reducing misunderstandings is also a crucial issue in cross-cultural understanding~\cite{Beech2022Mis,House1997Mis}. 
Taracenko\etal~\cite{Taracenko2018THE} suggest that fundamental educational training is necessary, such as enhancing basic knowledge of different countries, and learning about relevant languages and cultural characteristics. 
However, this process might be challenging and require significant effort. 
The development of LLMs provides a technical foundation for minimizing misunderstandings in cross-cultural understanding. 
Regarding data sources, Liu\etal~\cite{liu2023equitable} developed a Cross-Cultural Understanding Benchmark (CCUB) dataset aimed at achieving multicultural inclusivity and representation in generated images. 
In terms of exploration processes, NormSage~\cite{fung2022normsage} offers a framework for discovering dialogue-driven, multilingual, and multicultural norms, based on language model prompting and self-validation.
Inspired by these prompt engineering approaches, we explored the incorporation of visual elements in TCPs with LLMs to facilitate cross-cultural understanding.
We curate an expert-annotated TCP dataset based on the concept of \rw{cultural} norm to provide accurate information for LLM generation.

\subsection{Incorporating with Large Language Models}

After scaling the parameter capacity of pre-trained language models~\cite{kaplan2020scaling}, researchers found those LLMs showed surprising performance (called emergent abilities~\cite{wei2022emergent}) in solving diverse, sophisticated tasks~\cite{zhao2023survey,bubeck2023sparks,touvron2023llama,brown2020language}.
With their outstanding instruction-following~\cite{zhang2023instruction}, in-context learning~\cite{brown2020language,von2023transformers}, and reasoning~\cite{wei2022chain} abilities, LLMs have been leveraged to empower various domain applications.
These include a few attempts~\cite{xu2024ai,huang2023culturally} to apply LLMs on cultural and social sciences, which mainly focus on studying LLMs' inherent social bias~\cite{naous2023having,adilazuarda2024towards} and modeling the sociocultural norms~\cite{sky2023sociocultural,ramezani2023knowledge,rai2024cross}.

Unlike previous works, this paper is the first to adopt the powerful GPT-4~\cite{achiam2023gpt} model as the inference backend of our visual analytic system CultiVerse. 
We provide empirical evidence to show that LLMs' profound world knowledge can benefit cross-cultural understanding. 
Our carefully designed pipeline enhances the usability and authenticity of culture translation for the public interest.

\section{Design requirement analysis}
\label{sec:formative}
We have collaborated with domain experts to inform our system design, which aims to engage a broader audience with TCP through a cross-cultural dialogue. 
We interviewed four professors, two in TCP and two in cultural translation. E1 and E2 have 15 and 30 years of TCP research experience. E3 has dedicated 15 years to studying Chinese-Japanese literature and has a deep understanding of TCP. E4 focuses on comparative educational research between China and the UK.
None of them is a co-author.
Through semi-structured, one-on-one interviews, we explored a hypothetical scenario where experts assist non-natives in understanding a TCP. 
Then, we developed a prototype system incorporating ChatGPT-4 Turbo~\cite{chatgpt4turbo} to re-examine the scenario. 
All experts have recognized LLM's ability to bridge cultural and language barriers for cross-cultural understanding.
However, the inherent complexity of TCP's cultural connotations presents challenges that primitive LLM applications cannot adequately address. 
Through this iterative process, we collected expert feedback on strategies for promoting art appreciation and the challenges of using LLMs for cross-cultural understanding. 

\subsection{Painting analysis process}
\label{tcp_process}
E1 and E2 suggest three levels of comprehension (L1-L3) that progressively deepen an audience's appreciation of TCP, requiring increasing cultural understanding and artistic engagement.

\begin{figure}[t]
\centering
\includegraphics[width=\linewidth]{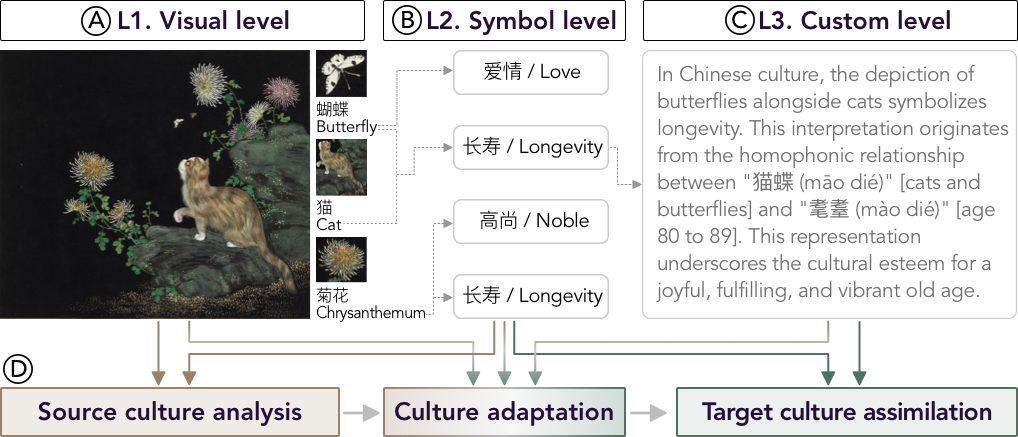}
\caption{The design requirements are inspired by TCP comprehension levels (A-C) and the literature translation process (D).}
\label{fig:requirement}
\end{figure}

\textbf{L1. Visual level}
focuses on the immediate observation and the surface appreciation of art's visual elements. For instance, Shen Zhenlin's ``\textit{Elderly in Spring Painting}'' (\cref{fig:requirement}A) vividly portrays a cat, butterflies, and chrysanthemums with delicate brushwork. This initial stage emphasizes aesthetics appreciation, including recognizing elements within the TCP and appreciating the drawing techniques used~\cite{fong1971understand}. 

\textbf{L2. Symbol level}
involves identifying relationships between elements and understanding why these elements are presented in specific ways. It explores the symbolic meanings behind the co-occurrence of certain elements~\cite{sun2021tcpsymbol,williams2012symbol}. For example, butterflies generally symbolize love in Chinese culture, but when depicted with cats, as in Shen Zhenlin's painting (\cref{fig:requirement}B), they suggest longevity. This interpretation originates from the homophonic play between 
``\begin{CJK}{UTF8}{gbsn}猫蝶~\textipa{m\={a}odi\'{e}}\end{CJK} [cats and butterfly]'' and ``\begin{CJK}{UTF8}{gbsn}耄耋~\textipa{m\`{a}odi\'{e}}\end{CJK} [being from 80 to 89 years old],'' 
illustrating symbolic associations embedded within the artwork.

\textbf{L3. Custom level}
seeks to understand the painter's intent and the cultural phenomena represented by the symbols~\cite{williams1976custom,amigo2015fitting}. 
This deepest level of appreciation requires a comprehensive understanding of the cultural context to connect the artwork to broader cultural practices and beliefs.
For example, presenting the painting named ``\textit{Elderly in Spring Painting}'' (\cref{fig:requirement}C) as a gift reflects societal values placed on a joyful, fulfilling, and vibrant old age due to respect and care for elders.

\subsection{Design requirement for cross-cultural understanding}
We gain inspiration from E3 and E4's experience in the literature translation process, which includes source culture analysis, culture adaptation, and target culture assimilation~\cite{bassnett2007trans}, for the design goals of cross-cultural understanding.
We further contextualize them with TCP comprehension levels and summarize the design requirements.

\noindent \textbf{G1. Source culture analysis.}
The foundation of accurate cultural translation lies in a deep understanding of the source culture.
E4 warns against the risks of biased translations stemming from insufficient understanding of the source material, akin to the notion ``garbage in, garbage out,'' which can significantly increase the chance of invalid LLMs' responses.
To mitigate this, LLMs should be provided with accurate information about the three levels of TCP comprehension, ensuring a more accurate and culturally sensitive translation process.

\textbf{R1. Identifying visual elements.}
From natural to mythological elements, TCPs are rich in artistic expressions and drawing techniques that differ from other artistic styles (\eg oblique projection~\cite{zhang2024tcp}). Developing models trained with diverse representative samples is crucial for accurate element identification at the visual level (L1).

\textbf{R2. Visualizing possible associations between elements.}
The tradition of replicating classics in TCPs (called ``to transmit models by drawing''~\cite{zhang2024tcp}) suggests that widely accepted cultural symbols and elements are repeatedly drawn.
Therefore, presenting recurring elements among different paintings can shed light on the potential connections within their original context~\cite{kamkwai2024prismatic}.
Moreover, transparently embedding authoritative historical and cultural resources in the system offers users and LLMs a trusted foundation for generating content with cultural insights~\cite{feng2023xnli}. 
These help understand the symbol level (L2).

\noindent \textbf{G2. Culture adaptation.}
Adapting elements and symbols from the source culture to fit the target culture's norms requires a delicate balancing effort, often challenged by disparities in symbolic and customary meanings. 
Gathering relevant information from both cultures is important to refine and adjust the adaptations. 
This iterative process enhances user engagement by offering familiarity and new insights.

\textbf{R3. Specifying the target culture.}
LLM responses can accommodate diverse user backgrounds.
To bridge language and cultural barriers, specifying users' cultural contexts and offering feedback on cultural relevance facilitates more meaningful and personalized translations.

\textbf{R4. Aligning cross-cultural symbols.}
The challenge of aligning symbolic meanings across cultures stems from L1-mismatches (where different elements represent the same symbol) and L2-mismatches (where the same element represents different symbols). The system should enable users to visually compare elements and symbols from both cultures to identify parallels or divergences in symbolic meaning.

\textbf{R5. Supporting multimodal understanding.}
E4 notes that illustrations and other visual aids are common strategies for literature translation.
They depict relatable scenes or realistic source culture objects, which are particularly useful for L3-mismatches where customs diverge significantly between cultures. Multimodal translations provide an enriched and engaging context for cross-cultural understanding.

\noindent \textbf{G3. Target culture assimilation.}
Integrating aligned symbols and customs into the target culture requires a process of creative exploration that honors cultural nuances. Employing LLMs to generate content rich in cultural relevance, such as narratives that find echo in both the source and target cultures, showcases a profound cross-cultural understanding. Nevertheless, such content should maintain cultural appropriateness.

\textbf{R6. Supporting creative exploration.}
By offering structured guidance around the users' natural curiosity, the system could spark a desire to explore TCPs across different cultural contexts. This curiosity-led approach facilitates users to form a personal connection with the materials. Encouraging questions and self-directed discovery allows for a richer, more immersive cross-cultural experience.

\textbf{R7. Validating norm acceptability.}
E3 and E4 found that some LLM-generated exploration results may deviate from the source culture's values. Since a general audience often lacks awareness of technical limitations such as content hallucination~\cite{zhang2023siren}, taking the result without validation can risk cultural biases. The system should have a safeguarding mechanism for validating the cultural appropriateness of content and preserving the integrity of cross-cultural understanding.

\section{Source \rw{cultural} norm dataset construction}
\label{sec:dataset}
This section describes the \textit{TCP \rw{Cultural} Norm Dataset (TCP-CND)} for source culture understanding (G1). 
We adopt a semi-automated approach for efficient, accurate dataset construction. 
Four TCP domain experts are invited to annotate TCP data, aiding in selection, optimization, and annotation. 
We select English as the primary language for the dataset to accommodate its wide adoption in the global population and LLMs' language ability, reducing the likelihood of misunderstandings in further cross-cultural translation processes.

\begin{figure*}[ht]
\centering
\includegraphics[width=1.0\linewidth]{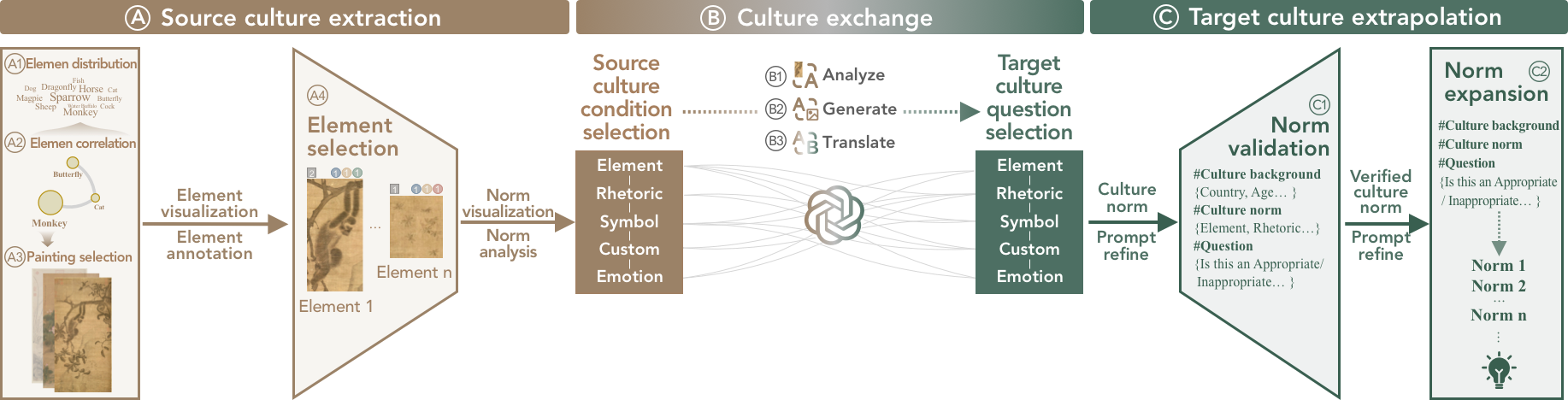}
\caption{The CultiVerse workflow comprises three steps: source culture extraction (A), culture exchange (B), and target culture extrapolation (C). Initially, user observe element distribution (A1), examine element correlations (A2), and select painting based on exploratory interests (A3), followed by choosing element for analysis (A4). Subsequently, user analyze (B1), generate (B2), and translate (B3) cultural norms in conjunction with LLMs. Finally, users validate (C1) and expand (C2) the translation results.}
\label{fig:workflow}
\vspace{-6pt}
\end{figure*}

\subsection{\rw{Cultural} norm definition}
\label{sec:feature_extraction}
Based on our requirement analysis and cultural studies~\cite{sent2022commemorating}, we define the data structure of \textit{\rw{Cultural} Norm (CN)} as follows:
\begin{equation}
  \label{cn}
  CN = (Element-Rhetoric-Symbol-Custom-Emotion)
\end{equation}
where \textit{Element} refers to the objects that appear in the painting. 
\textit{Rhetoric} denotes the technique used to create a symbolic association between the ``element'' and the ``symbol.''
\textit{Symbol} refers to the symbolic meaning that the ``element'' is associated with, using a specific ``rhetoric`` technique. 
\textit{Custom} corresponds to the cultural, societal, or traditional practice or belief associated with the element and its symbol. 
\textit{Emotion} describes the feeling or sentiment associated with or evoked by using the symbol within the specific cultural context.

\subsection{Data collection and annotation}
\label{dataset}
We collect TCP information from two open datasets, the National Palace Museum (NPM)~\cite{NPM} and the Chinese Iconography Thesaurus (CIT)~\cite{CIT}. 
Each painting contains the following data information:

\textit{Painting image and background information.} 
Under the guidance of experts, we carefully selected a collection of paintings, such as flower and bird paintings, imbued with diverse implications that are more conducive to public users. We excluded landscape paintings and portraits with complicated historical backgrounds. 
Ultimately, we obtained 534 pieces of paintings. Each painting is characterized by the artist, dynasty, medium, dimensions, and location.

\textit{Painting element.} 
The CIT dataset establishes a thesaurus for TCP, including general elements (\eg plants, animals, characters, and events) and proper nouns (\eg locations, myths, and literature).
However, some elements, such as ``looking upward (bird)'' representing animal forms, offer limited assistance in cross-cultural understanding. 
We collaborated with experts to filter these elements. Finally, we selected a total of 226 elements (plant 94, animal 86, fruit 16, other 13, composite elements 17).
However, the lack of instance-level annotations (\ie bounding boxes and categories) hinder the in-depth exploration of the painting. We adopted a human-in-the-loop labeling pipeline to guarantee high data quality at reasonable costs.
We utilized GroundingDINO~\cite{liu2023grounding}, a prompt-based object detection model, to obtain the bounding boxes of visual elements.
Due to natural degradation, some TCPs exhibit color fading~\cite{tang2023pcolorizor} and contain elements in unique art forms, undermining automatic recognition's reliability and fall short of our annotation standards.
We invite four experts to refine and augment the annotations. 
Each annotation is thoroughly verified by two experts for accuracy and consistency in (\textit{Image $-$ Element}) alignment (R1).

\subsection{Norm extraction}
We cooperate with experts to extract, organize, optimize, and enrich norms from the ``Chinese Cultural Symbol Dictionary''~\cite{CCSD}. Ultimately, we obtained 505 cultural norms.

\textbf{Element, symbol, and custom extraction.}
\label{sec:element_def}
The dictionary contains various symbolic meanings of the elements and their origins in Chinese culture.
We classify elements into \textit{atomic} and \textit{composite elements}. 
The composite elements are generated from multiple atomic elements via the \textit{and} logical operator to represent more complicated contexts. 
For instance,
atomic~(\begin{CJK}{UTF8}{gbsn}蜂~\textipa{f\={e}ng}\end{CJK} [Bee])
~\&~
atomic~(\begin{CJK}{UTF8}{gbsn}猴~\textipa{h\'{o}u}\end{CJK} [Monkey])
indicates the 
{composite~(\begin{CJK}{UTF8}{gbsn}蜂猴~\textipa{f\={e}ngh\'{o}u}\end{CJK}
=
\begin{CJK}{UTF8}{gbsn}封侯~\textipa{f\={e}ngh\'{o}u}\end{CJK} [being ennobled as a marquess]).
However, not all of these apply to our study. For instance, elements such as ``cold,'' ``constellations,'' ``dust,'' and ``earth'' pose challenges in mapping their visual representation through painting. As such, we refine these data and ultimately extract 1453 elements, along with their various symbols and customs (R2).

\textbf{Rhetoric and emotion identification.} 
To foster a more profound comprehension of the emotions associated with customs, experts annotate each ``element - symbol - custom'' pair. The emotions can be categorized into three types: ``positive,'' ``negative,'' or ``neutral.''
Experts also annotate the rhetorical techniques employed in the ``element - symbol'' pairs, elucidating the method of transforming an ``element'' into a ``symbol, '' such as
(\begin{CJK}{UTF8}{gbsn}蜂猴~\textipa{f\={e}ngh\'{o}u}\end{CJK} 
- Homophony -
\begin{CJK}{UTF8}{gbsn}封侯~\textipa{f\={e}ngh\'{o}u}\end{CJK}).
We summarize six rhetorical techniques from the paintings and iterate their definitions with E1 and E2. The definitions for each rhetorical technique are as follows:

\raisebox{-0.3ex}{\tikz\fill[Iconic] (0,0) circle (0.8ex); }~
Iconic: the form of the sign is connected to its meaning.

\raisebox{-0.3ex}{\tikz\fill[Homophony] (0,0) circle (0.8ex); }~
Homophony: two words have the same pronunciation but different meanings, origins, or spelling.

\raisebox{-0.3ex}{\tikz\fill[Homophonic pun] (0,0) circle (0.8ex); }~
Homophonic pun: a type of pun that exploits the fact that words sound similar but have different meanings.

\raisebox{-0.3ex}{\tikz\fill[Synonym] (0,0) circle (0.8ex); }~
Synonym: a word that means exactly or nearly the same as another word.

\raisebox{-0.3ex}{\tikz\fill[Homograph] (0,0) circle (0.8ex); }~
Homograph: a word that shares the same form as another word but has a different meaning.

\raisebox{-0.3ex}{\tikz\fill[Satire] (0,0) circle (0.8ex); }~
Satire: a word with a satirical meaning often reflects negative aspects of society.

\section{CultiVerse}
\label{sec:cultiverse}

This section describes the workflow, engineered prompts for the components, and interactions between views in CultiVerse (\cref{fig:workflow}).

\subsection{User background setting}
\label{background}
Initially, users are required to configure background information (G2, R3).
This role information will be utilized as a prompt setting, transmitted to the LLM, and integrated throughout the entire analysis workflow. The data structure of \textit{User Background (UB)} is as follows:
\vspace{-2pt}
\begin{equation}
  UB = (Country,~Age,~Education,~FWC,~FWTCP,~Note)
  \label{ub}
  \vspace{-2pt}
\end{equation}
where \textit{FWC} and \textit{FWTCP} refer to the user's familiarity with Chinese culture and TCP on a scale of 1 to 5, respectively.
\textit{Note} allows users to specify contextual information, such as cultural preferences for anime.

\subsection{Source culture extraction
component}
\label{extract}
The source culture extraction component (\cref{fig:workflow}A) offers an element feature view that illustrates data features based on the TCP-CND (G1, R1, R2). 
Users can then explore these data features and select a painting that interests them. The chosen element from the painting is subsequently forwarded to the following analysis stages.

\begin{figure}[t]
\centering
\includegraphics[width=\linewidth]{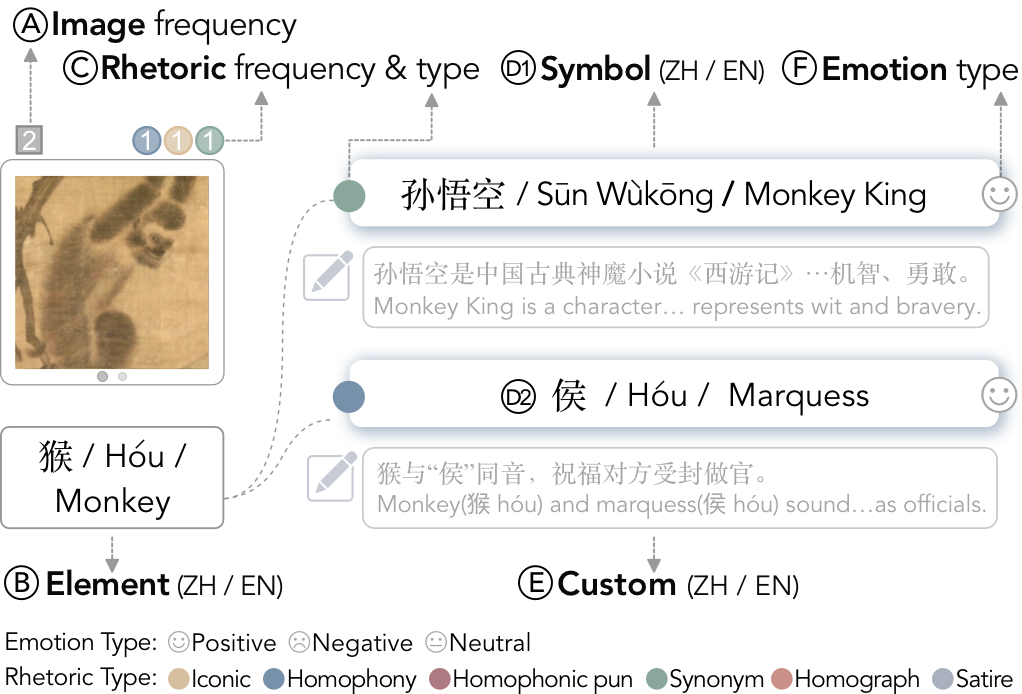}
\caption{The element selection view highlights the frequency of elements in the dataset (A) and counts of other factors in the \rw{cultural} norm (B-F).}
\label{fig:nd}
\vspace{-6pt}
\end{figure}

\subsubsection{Element feature view}
\label{feature}
The element feature view visualizes all identified elements derived from the dataset (R1). To enhance the understanding of data features, this view illustrates each element's frequency, relevance, and overall distribution across various categories (R2). We encode the elements through distinct channels to differentiate them.

\textbf{Visualizing elements in colors, sizes, and lines.}
The extracted features could be either atomic or composite elements (see \cref{sec:element_def}).
We use different colors to distinguish the four categories of atomic elements (\ie \eventtype{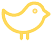}{Animal}, \eventtype{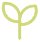}{Plant}, \eventtype{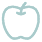}{Fruit}, and \eventtype{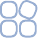}{Others}). 
We encode the number of elements by size to emphasize the data distinction (\cref{fig:workflow}A1).
Zooming in an atomic element reveals its composite elements.
The composite elements combine the lines of their corresponding elements. The lines' thickness encodes the number of element's relevance (\cref{fig:workflow}A2).

\subsubsection{Painting selection view}
\label{detail}
When a user selects an element in the element feature view, a related painting is displayed in the painting selection view (\cref{fig:workflow}A3). 
The user can then choose one for further analysis. Element annotations from the TCP-CND, which includes the element bounding boxes and their corresponding labels, are shown in the painting (R1). 
If there are elements of interest to the user that have not been annotated, we provide a manual annotation tool to assist the user in this task. 
The newly added annotations are also recorded in the element selection view (\cref{fig:workflow}A4). 
The painting's background information is presented on demand for more detailed information.
Image interaction tools (\eg fullscreen, zoom in, zoom out) are provided to facilitate close investigation of visual elements in TCPs.

\subsubsection{Element selection view}
\label{selection}
The element selection view (\cref{fig:teaser}A4) provides detailed information about the elements of the selected painting (R1). 
Numbers in \cref{fig:nd}A represent the frequency of each image.
The related elements of the image are displayed below (\cref{fig:nd}B) in Chinese and English.
The colors in \cref{fig:nd}C indicate the categories of rhetorical types involved with each element, and the numbers denote the frequency of usage of these types. 
When composite elements are present, they are linked to related atomic elements and enhanced with icons for clearer representation (R2).
The elements are responsive to user-annotated elements if they belong to the dataset, facilitating ad-hoc and comprehensive exploration.

\subsection{Culture exchange component}
\label{adapt}
Upon selecting an element, users can explore its diverse symbolic meanings in this component by harnessing the capabilities of the LLM (\cref{fig:workflow}B).
The selected \rw{cultural} norm can be adapted to the target culture for cross-cultural understanding (\cref{fig:workflow}B1-B3) (R4-R6).

\subsubsection{Source culture exploration view}
\label{explore}
The source culture exploration view (\cref{fig:teaser}B1) visualizes the source \rw{cultural} norm and provides multiple ways to explore it (G1).
When an element is selected, all related norms are displayed. \cref{fig:nd}D1 and D2 present the different symbols in both Chinese and English. The left side illustrates the rhetoric employed in the `element-symbol' pairs, differentiated by various colors. Beneath each symbol is a custom explanation, which elucidates why the element has been assigned this symbol, presented in both Chinese and English (\cref{fig:nd}E).  
The right side of the symbol indicates the emotion associated with its use within a specific culture, categorized into three types and distinguished by different icons (\cref{fig:nd}F).

\textbf{Explore with question-answering.} We utilize the powerful question-answering capabilities of the LLM to allow users to pose queries about any unclear aspects of the source \rw{cultural} norm. To provide guidance and prompts, we have pre-set three recommended questions. Additionally, users are free to ask their questions. The structure of the question-answering prompt is as follows:
\vspace{-2pt}
\begin{equation}
  \label{eq_qa}
  QA = LLM(UB,~CN,~PR~|~FQ)
\vspace{-2pt}
\end{equation}
Where \textit{UB} denotes the user background in \cref{ub}, \textit{CN} represents the source \rw{cultural} norm in \cref{cn}, \textit{PR} denotes the prompt recommendation, and \textit{FQ} signifies the free question. 
The question-answering process incorporates a memory function, enabling users to continue asking questions based on previous responses. A deletion function is also provided on the right side of the dialogue box, allowing users to remove unnecessary dialogue boxes to conserve viewing space.

\textbf{Explore with image generation.}
To enhance users' understanding of the source \rw{cultural} norms, we incorporate LLM into image generation (IG). 
It aims to augment the comprehension of source \rw{cultural} norms visually (R5). To enable the LLM to generate images that align with users' cognition, we designed the following prompt:
\vspace{-2pt}
\begin{equation}
  \label{eq_ig}
  IG = LLM(UB,~CN,~T)
\vspace{-2pt}
\end{equation}
where \textit{T} corresponds to the image generation task. 
Once the image is generated, options for deletion and regeneration are provided to cater to users' requirements for multiple generations.

\subsubsection{Culture transfer view}
\label{transfer}
The culture transfer view (\cref{fig:teaser}B4) integrates the LLM to offer culture translation services to users (R6). 
The output of LLM is visualized with the same design scheme as the source \rw{cultural} norm (\cref{fig:nd}) to maintain visual coherence. 
The view also supports Image Generation and Question-Answering to assist users in further exploration.

\textbf{Multiple conditions.} To cater to the diverse culture transfer needs of users, we provide an interactive function, enabling users to freely choose a part or all of the source \rw{cultural} norm as the conditions (C):
\vspace{-2pt}
\begin{equation}
\fontsize{8.5pt}{12pt}\selectfont
  \label{eq_c}
  C \in \textit{source culture} \{\textit{Element}, \textit{Rhetoric}, \textit{Symbol}, \textit{Custom}, \textit{Emotion}\}
\vspace{-2pt}
\end{equation}
For question (Q) formulation, users can interactively choose whether on a single part or multiple parts of the target \rw{cultural} norm:
\vspace{-2pt}
\begin{equation}
  \fontsize{8.5pt}{12pt}\selectfont
  \label{eq_q}
  Q \in \textit{target culture} \{\textit{Element}, \textit{Rhetoric}, \textit{Symbol}, \textit{Custom}, \textit{Emotion}\}
\vspace{-2pt}
\end{equation}
The selected conditions and questions become a part of the prompt:
\vspace{-2pt}
\begin{equation}
  \label{eq_ct}
  CT = LLM(UB,~CN,~E,~C,~Q,~OF)
\vspace{-2pt}
\end{equation}
where \textit{E} refers to the explanation of the definition, such as rhetoric types. 
\textit{OF} refers to the pre-defined output format for the view.

\subsection{Target culture extrapolation view}
\label{assimilate}
In this stage, users can leverage LLM to verify the appropriateness of the explored norms and expand on similar norms (\cref{fig:teaser}D) (G3, R7).

\subsubsection{\rw{Cultural} norm verification}
We put an ``Appropriate'' button, enabling users to obtain a question template to validate cultural appropriateness (R7).
The content contained in the template is as described in the previous paragraph. 
After sending the question to LLM, it can give a judgment based on the target culture setting and current translation path, and analysis about it.

\subsubsection{\rw{Cultural} norm expansion}
Also, we provide an ``Inference'' button for extending similar norms in a way beyond the workflow (R6), drawing inferences from one instance as a looser and freer supplement to the previous translation method.

\section{Evaluation}
\label{sec:eva}

We conducted a user study to evaluate how CultiVerse assists a broader audience in achieving cross-cultural understanding through TCP. 
We invited ten participants (P1-P10) from diverse cultural backgrounds to assess the effectiveness and usability of CultiVerse through comparative studies and free exploration.
\rc{The explorations by P3 and P10 are presented as the case studies, showing how CultiVerse facilitates cross-cultural understanding when users have knowledge in source culture (\cref{sec:case1}) and target culture (\cref{sec:case2}).}
We use labels (\ding{172}-\ding{181}) to denote the new insights gained during their exploration.
We summarize user feedback on the system design, workflow, and exploration patterns.

\begin{figure}[t]
\centering
\includegraphics[width=\linewidth]{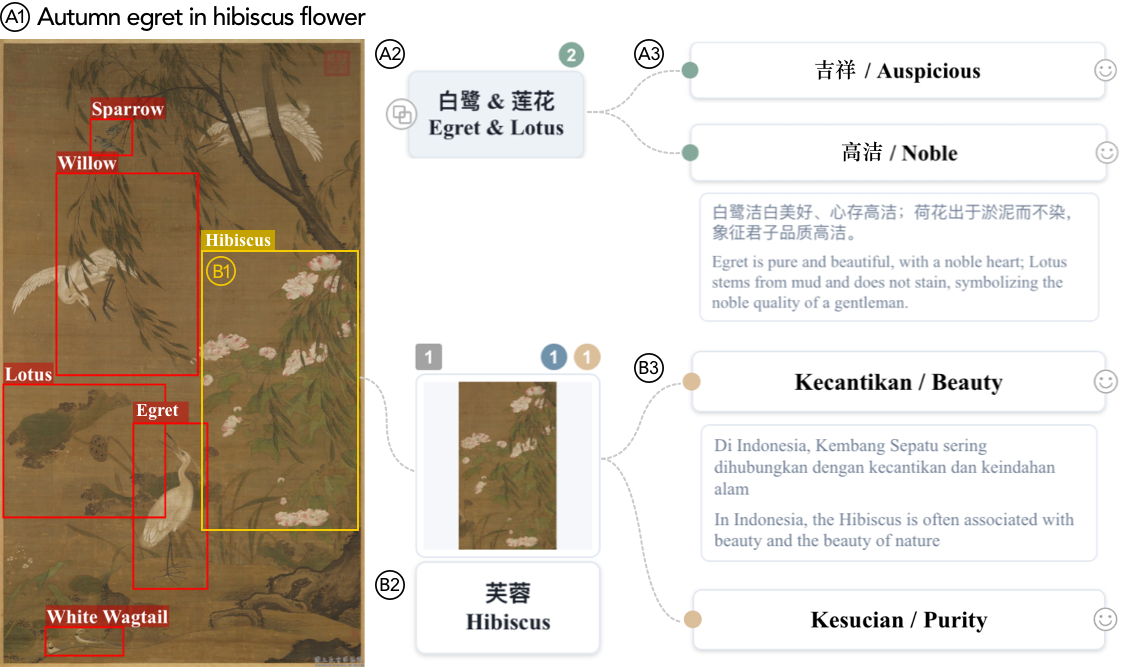}
\caption{Case 2 explores (A1) ``\textit{Autumn Egret in Hibiscus Flower Landscape}.'' (A2) Analyzing compositional elements to (A3) understand their significance in Chinese culture; and (B1) using interactive annotations to (B2) explore new elements (B3) in Indonesian culture.}
\label{fig:case2_1}
\vspace{-6pt}
\end{figure}

\begin{figure*}[t]
\centering
\includegraphics[width=1.0\linewidth]{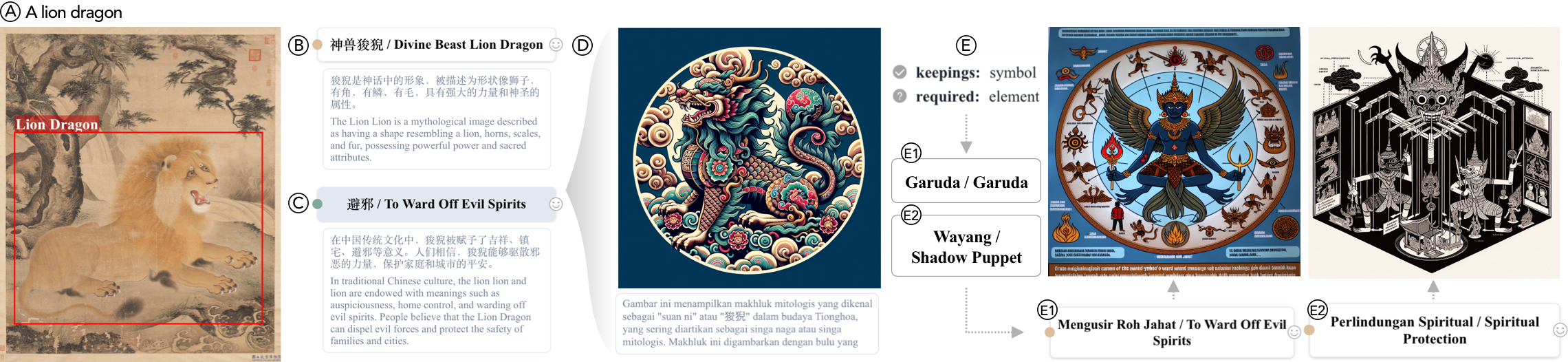}
\caption{In the second exploration phase, an unfamiliar element (A) was analyzed to understand its symbols in Chinese culture (B, C), with image generation enhancing visual understanding (D). The second symbol was then translated into the Indonesian cultural context (E), further improving comprehension through image generation (E1,E2).}
\label{fig:case2_2}
\vspace{-0.2in}
\end{figure*}

\subsection{Exploring diverse cultures}
\label{sec:case1}
P3 (nicknamed Kay) is from China and, working as a consultant, has a deep appreciation for Japanese culture. She holds a Japanese-Language Proficiency Test N2 certificate and has traveled to Japan numerous times. In the target culture panel, she sets her sights on Japan, embarking on an in-depth exploration and study.

\textbf{Source culture extraction.}
Initially, Kay closely looked at the distribution of the \textit{element feature view} (\cref{fig:teaser}A1). \ding{172} Among the word cloud, ``monkey'' popped out and interested her, reminding her of her experiences observing monkeys up close at the Jigokudani Monkey Park in Japan. This triggered her curiosity about monkeys' representation and potential symbolism within TCPs.

The \textit{painting selection view} (\cref{fig:teaser}A2) displayed all eight paintings from the dataset that included the monkey element.
\ding{173} Kay selected and obtained background information about the ``Monkey Painting'' by Yan Hui in the Mongol Empire (1271-1368) (G1). 
The two monkeys and a swarm of bees were automatically identified in the painting (L1, R1).
\ding{174} She marveled at the intricate depiction of bees, stating that she would not have noticed them in the painting without the annotation. 
\rc{She enlarged the painting for a close examination.
She thought the ``leaf'' might be a missed element, and annotated it herself (\cref{fig:teaser}A3).}

\rw{Kay navigated to the \textit{element selection view} (\cref{fig:teaser}A4) for the elements' features. Through the statistics in the top-right corner, \ding{175} she noticed that the monkey is associated with three symbols and rhetorical techniques (L2), and forms a composite element (R2) with the bees, termed ``bee \& monkey.'' \ding{176} The newly annotated element ``leaf'' did not carry symbolic meanings.
She clicked on the ``bee \& monkey'' for further analysis in the \textit{culture exchange component} (\cref{fig:teaser}B).}

\textbf{Culture exchange.}
\rw{In the \textit{source culture exploration view} (\cref{fig:teaser}B1), ``bee \& monkey'' was explained as symbolizing ``being ennobled as a marquess'' because the two phrases are associated with homophony. They frequently appear together in paintings, embodying aspirations for upward mobility~\cite{wang2021interactive} (\cref{fig:teaser}B2). 
\ding{177} Kay was surprised about the implications of ``bee \& monkey.''
She wondered if other cultural contexts exist for the connection between the monkey and the marquess (R6). 
While the typical cultural research could take pages of reading, she simply conversed with the LLM and got a satisfactory result. 
\ding{178} Monkeys are symbols of intelligence and wit; Therefore, drawing one may imply a tribute to the wisdom and mastermind of nobles (L3).
Also, she prompted the LLM to generate an image related to an ancient Chinese marquess (\cref{fig:teaser}B3). 
This visual aid spelled out the hidden meaning in the painting for her (R5).}

Kay wondered what elements in Japanese culture could represent a ``marquess.'' She selected ``symbol'' as the condition in \cref{fig:teaser}B1, and ``element'' as the question in \cref{fig:teaser}B4 (R4). Upon clicking the translation button, \ding{179} the LLM provided two symbols, namely ``chrysanthemum'' and ``family crest.'' 
These results aligned with her understanding of Japanese culture. 
The explanations provided for each symbol facilitated a deeper understanding for Kay. 
\ding{180} Upon clicking the image generation button, the LLM produced an image aligned with Japanese visual aesthetics (\cref{fig:teaser}B5) (R5).
Consequently, Kay gained insight into the differences and underlying reasons for how ``marquess'' is represented through elements in Chinese and Japanese cultures (G2).

\textbf{Culture extrapolation.}
\rw{Kay wanted to know more about symbols representing ``marquess'' in other countries. In the \textit{target culture extrapolation view} (\cref{fig:teaser}C), she clicked the inference button to get a series of recommended symbols associated with marquess from different cultures (R6). \ding{181} For example, the British culture symbolizes a marquess with a coronet mounted with pearls, while Indian and French cultures often associate it with the tiger and the fleur-de-lis.}

\subsection{Exploring a mysterious TCP}
\label{sec:case2}
P10 (nicknamed Lily) is an undergraduate student, born and raised in Indonesia, who is strongly interested in Chinese culture but not yet familiar with TCP. This case is highlighted in the supplementary video.

\textbf{Set target culture information.}
First, Lily configured the \textit{target culture setting view} to reflect her cultural background (R3), Indonesia. She then evaluated her familiarity with Chinese culture and TCP, assigning ratings of 3 and 1, respectively. After establishing these parameters, she began to explore the system.

\textbf{Explore composite elements.}
The lotus is Lily's favorite plant. 
\rw{\ding{172} She found the lotus and the egret were closely related in the \textit{element feature view} (R2).
She picked the ``autumn egret in hibiscus flower landscape'' (\cref{fig:case2_1}A1) from the view. 
\ding{173} The other two views in the component provided data distribution of the elements and their symbolic meanings.}
Furthermore, \ding{174} she discovered that the lotus and egret form a symbolic combination in Chinese culture (\cref{fig:case2_1}A2), representing ``auspiciousness'' and ``nobility,'' respectively. She continued to learn about the cultural significance of this composite element through the \textit{source culture exploration view}. \ding{175} The combination symbolizes ``nobility'' because the egret, with its pure white beauty and noble heart, and the lotus, rising in purity from the mud, embody the esteemed quality in Chinese culture (\cref{fig:case2_1}A3) (G1).

\textbf{Label new elements.}
\rw{The painting's title contains ``hibiscus flower,'' but it was not annotated automatically.}
\ding{176} Lily examined and eventually identified the hibiscus within the painting (\cref{fig:case2_1}B1) (R1). Then, her interest was directed to its cultural significance (\cref{fig:case2_1}B2). 
\rw{In the \textit{culture exchange component}, she translated the ``element'' from the source culture to the ``symbol'' in the target culture.}
\ding{177} She discovered through LLMs that in Indonesian culture, the hibiscus flower symbolizes ``kecantikan [beauty]'' and ``kesucian [purity]'' (\cref{fig:case2_1}B3). 
This enhanced her understanding and appreciation of her culture (G2).

\textbf{``What is a lion dragon?''}
When catching a glimpse of ``lion dragon'' in the \textit{word cloud view}, Lily felt extremely curious. ``\textit{What is a `lion dragon?' Both lion and dragon, or neither of them?}''
Then, she selected a painting containing the ``lion dragon'' (\cref{fig:case2_2}A). By carefully observing the painting and conversing for LLM-generated contents in texts and images (\cref{fig:case2_2}D), \ding{178} she roughly understood the mythological creature ``lion dragon'' (G1, R5).
\ding{179} She noticed that it is associated with the symbol ``to ward off evil spirits'' (\cref{fig:case2_2}C). 
\ding{180} Inspired by this, she wondered what elements in Indonesia represent the same meaning (\cref{fig:case2_2}E) and successfully got the result: ``garuda''  and ``wayang [shadow puppet]'' with their explanations (\cref{fig:case2_2}E1, E2) (R4).

\textbf{Validate norm acceptability.}
Her knowledge of these two elements was limited to a few traditional stories she remembered. 
Consequently, she chose one of the findings and clicked the ``appropriate'' button in the \textit{target culture extrapolate view} to validate it (R7). 
The system generated a prompt that included all the information she needed. After submitting it to LLM, \ding{181} she received highly positive feedback, completed with detailed reasons and additional recommendations. 
Drawing on the traditional stories she recalled and the explanations provided by the system, she concluded that this adaptation was acceptable (G3).

Finally, within a 30-minute exploration period, she explored four TCPs and translated two cultural contexts. She was excited that the system offered her a new perspective on TCP and enabled her to discover fascinating insights from other cultures (R6).

\subsection{User study with diverse backgrounds}

\textbf{Participants.}
We recruited ten participants (P1-P10) from eight countries, striving for a balanced distribution across geographical and ethnic origins. Their cultural backgrounds are described in \cref{fig:freeex}. 
None of the participants had prior experience with the system. Each participant signed written consent and received a compensation of \$13.80.

\textbf{Comparative study design.}
\rc{This study adopted a within-subject design with balanced cross-over conditions to counter-balance and control for learning effects.
The ``\textit{Three Friends of Winter}''~\cite{PA} and ``\textit{Apricot Blossom and Peacock}''~\cite{PB} were chosen as the subjects.
Participants were asked to explore these paintings serendipitously under different conditions and seek cultural insights.
They are split into two groups: P1-P5 explored the first painting under the baseline condition and the second with CultiVerse, and vice versa for P6-P10.
The order of conditions was randomized.}

\textbf{Baseline condition.}
\rc{The baseline simulates non-natives' typical encounters with TCPs (\eg museum visits and online media). To parallel the \textit{source culture extraction} component in CultiVerse, participants watched a 2-minute expert commentary on the selected painting. 
The commentaries were created by an expert who annotated the \textit{TCP-CND} and instructed to explain the paintings to those unfamiliar with Chinese culture.
Following the video, participants can access untailored LLMs (same models as CultiVerse without designed prompts) and web search engines to support subsequent exploration.}

\textbf{Procedures.}
\rw{Each study session lasted 90 minutes. 
It began with a 10-minute introduction to the study background and collection of participants' demographic information and consent.
Then, they started the first assigned condition for 10 minutes, followed by a 5-minute feedback session. Participants answered semi-structured interview questions and completed a five-point Likert-scale questionnaire based on the user engagement evaluation theory~\cite{o2010eva}.
The second condition and feedback session took another 15 minutes.
CultiVerse was introduced with an example-based introduction~\cite{yang2023examples} before the start of that condition.
After experiencing both conditions, participants engaged in a 30-minute open-ended exploration session with CultiVerse.
The session finished with a final debriefing to collect comprehensive feedback.}

\begin{figure}[t]
\centering
\includegraphics[width=\linewidth]{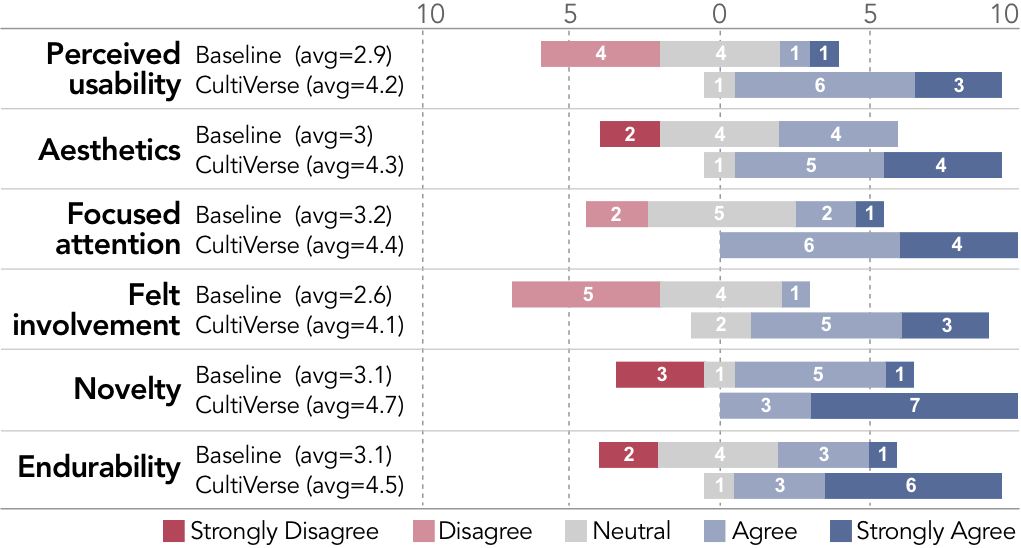}
\caption{The engagement-based questionnaire results regarding the effectiveness and usability of baseline and CultiVerse.}
\label{fig:eva}
\vspace{-6pt}
\end{figure}

\begin{figure}[t]
\centering
\includegraphics[width=\linewidth]{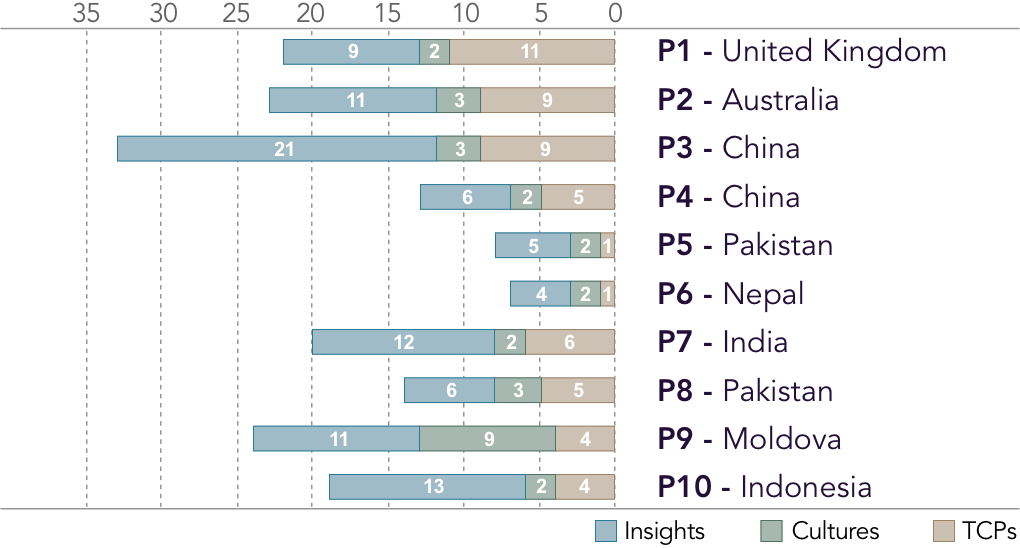}
\caption{The number of TCP that participants explored, the number of target culture explored and the number of insights during the open-ended exploration. The cultural backgrounds of participants are listed.}
\label{fig:freeex}
\vspace{-6pt}
\end{figure}

\subsection{Results}
\rw{The assessment for CultiVerse is consistently positive across all measured aspects, as shown in \cref{fig:eva}.
Detailed findings are outlined below.}

\textbf{Perceived usability}
assesses whether users can successfully and smoothly perform the desired tasks. 
All users reported that CultiVerse's analytical flow was clear and straightforward for enhanced exploration efficiency. 
P9 commented, \textit{``The system compares Chinese culture with the target culture simply and intuitively.''}
P2 also remarked on the rapid exploration capabilities, stating, \textit{``I reviewed 15 paintings and their symbolic meanings within 30 minutes, which gave me a good understanding of TCP.''} 
\rc{This efficiency contrasts starkly with the baseline, where participants found it challenging to articulate their inquiries on an unfamiliar topic.
The well-designed prompts in CultiVerse evidently enriched the dialogue between users and the system.}

\textbf{Aesthetics} 
evaluates users' impressions of the system's visual and interaction design. 
\rc{Positive feedback, primarily from P1, P3-P6, P8, and P9, concentrated on the \textit{source culture extraction} component.
P8 complimented the neat arrangement of paintings and objects, which provided visual comfort.
P3 particularly appreciated the \textit{element feature view}'s ability to ``\textit{intuitively understand the multi-layered meanings of the composite objects.}'' 
These designs effectively facilitated the comparison of cultural differences and similarities.
However, some feedback indicated that the LLM-generated images lacked authenticity compared to those from search engines. 
Meeting aesthetic expectations sometimes required multiple image regenerations and prompt engineering, particularly for less-known concepts.}

\textbf{Focused attention}
assesses users' perception of time passage and concentration during the exploration. 
All participants have positive ratings on this dimension and have shown enhanced usage efficiency.
\rc{For instance, P2 highlighted the integration of LLM functionality in the \textit{cultural exchange view}, which facilitated context-based dialogue and helped focus on elaborating specific cultural norms for deeper understanding. 
Conversely, the need for frequent scrolling back to previous conversations in the baseline condition was seen as a distraction from maintaining focus on the artwork.}
P10 stated, \textit{``Compared to traditional methods that require frequent switching between web pages, CultiVerse allows me to complete all explorations smoothly.''}

\textbf{Felt involvement}
reflects users' evaluation of the overall exploration experience.
\rc{In the baseline, participants used direct keywords from the commentaries (\eg painter's name) to guide their searches and dialogues with LLMs, leading to a uniform set of findings.
They tended to end exploration once they fully understood the commentary rather than further exploring based on personal interests.
In contrast, CultiVerse users reported a heightened sense of engagement due to the opportunity to expand their knowledge boundaries beyond the initial information (P2, P3, and P7-P10).
P10 described this cultural exchange process as addictive, similar to ``\textit{completing game tasks},'' due to the thrill of discovering new insights through experimenting with various combinations.
However, P1, P4, and P9 pointed out that selecting different cultural norms in the \textit{cultural exchange} component posed a steep learning curve, impacting their overall experience.}

\textbf{Novelty}
evaluates if users can discover surprising information. 
\rw{All participants reported gaining more cross-cultural insights than they did with the baseline.}
\rc{For instance, P7 discovered that crabs symbolize aggressive and overbearing behaviors through their unique sideways movement.
These nuanced cultural symbols are often inadequately translated or explained on the Internet.
Although LLMs in the baseline provided some support, users struggled with prompt refinements to get the desired results.
In contrast, CultiVerse assembles effective tools and customized prompts that facilitate richer cultural exchanges.
The explicit option to visualize concepts, such as the tailored representation of crabs moving sideways, has proven effective in helping users understand and remember these cultural symbols.}

\textbf{Endurability}
refers to the likelihood of users revisiting and recommending the system to others. 
P10 mentioned considering CultiVerse's information as a reference when choosing gifts for Chinese friends.
However, P2 observed that the LLM-generated content had stereotypical impressions, such as associating American culture with eagles and Australian culture closely with kangaroos.
This led to doubts about the system's value and intentions for future usage.

\subsection{Open-ended exploration}
We reviewed users' exploration data and documented three key variables: the number of explored TCPs, selected target cultures, and reported insights during the free exploration (\cref{fig:freeex}).

\rw{Unexpectedly, P4-P6 explored fewer TCPs and cultures and reported fewer insights than others. 
Although they have closer cultural distances~\cite{hofstede2011dimensionalizing} to China than those from European and Asia-Pacific countries (\eg P1 and P2), their outcomes were not as anticipated.
We identified the following reasons from semi-structured interviews.
First, the limitations of LLMs in processing information from lesser-dominant cultures led to inaccurate results and long wait times, reducing user trust and enthusiasm with CultiVerse.
For example, P4 reported fewer insights since P4 spent more time transitioning between the Han and Tibetan cultures, which was ineffective because LLMs made numerous incomprehensible errors in Tibetan languages.
Similar issues were observed with other less commonly represented languages like Pakistani (used by P5), negatively affecting the user experience. 
While the results can be translated into English, the translation loss and varied English proficiency of the participants may hinder the depth of understanding possible.
Second, participants may prefer an in-depth investigation over broader topics, leading to fewer insights in numbers. 
P6 dedicated significant time to composing poetry in Nepali for a TCP. 
P5 focused on reading custom information in the TCP-CND dataset.
This resulted in fewer explored paintings, cultures, and reported insights.}

\rw{As P3 belongs to the source culture in CultiVerse, P3 was expected to gain more insights than others.
However, P3's skilled use of LLMs also improved the result.
Unlike other participants, P3 relied much less on the TCP-CND dataset.
Instead, P3 frequently utilized the LLM's generative imaging and prompting to aid cross-cultural understanding. 
Due to her familiarity with the source culture, more time was spent on the \textit{culture exchange} and \textit{target culture extrapolation} components. 
P3 noted, ``\textit{Exploring whether known elements also hold special meanings in other countries helps me understand cultural diversity.}''
Similarly, P9 explored more cultures by translating the same elements into nine different cultures. 
P9 believed that doing so could facilitate deeper insights into subtle cultural differences.}

\section{Discussion}
\rw{We invited four domain experts (E3, E5-E7) with extensive experience in communication studies (E5), TCP education (E7), and cross-cultural translation (E3 and E6) for professional assessments. 
In this section, we conclude and discuss the lessons learned and the study's limitations.}

\rw{\textbf{Enabling new usage scenarios like the Swiss army knife.}
All four experts unanimously agreed on the significant potential of the project. 
Their diverse backgrounds contributed to a broad spectrum of suggested applications, further highlighting CultiVerse's versatility across different professional scenarios akin to the Swiss army knife.}

\rc{E3 praised CultiVerse for its suitability in classroom environments, particularly for its support of interactive and digital teaching methods that cater to various learning styles. 
The system promotes a pedagogical shift from traditional teaching methods to a more constructivist approach, encouraging active learning and knowledge construction.
E3 and E6 also recognized its effectiveness in introductory courses to reduce entry barriers in cross-cultural topics.
Nonetheless, they expressed concerns about LLM's ability to self-validate and provide justified information in higher-level academic settings.}

\rc{E5 appraised CultiVerse as an invaluable tool for overcoming cultural barriers in communication.
It can aid art exhibition tour guides deepen their understanding of artworks and enhance their presentation skills.
By translating cultural elements and norms to match the audience's background, tour guides can tailor their explanations to be more relevant and engaging.
P3's experience supports this, showing that users with background knowledge can utilize CultiVerse more effectively and efficiently.
This suggests that professionals such as tour guides and artists could particularly benefit from its features.}

\rc{E5 and E7 recognized CultiVerse's usability in supporting creativity and research, especially for ``cultural comparison'' studies.
Based on traditional paintings, CultiVerse can provide motivation and stimulate innovative ideas centered around historical and cultural themes.
This efficiency in generating ideas reduces the time designers spend brainstorming.
It allows them to devote more energy to the creation process, thereby enriching the variety and quality of their designs.}

\textbf{Generalizability and transferability.}
\rc{Our system exemplifies how existing cultural studies frameworks can be adapted to meet the future requirements of intelligent systems focusing on cross-cultural understanding.
We have discovered that personal curiosity is crucial in fostering such understanding, and CultiVerse has effectively stimulated this curiosity among users.
E3 noted that the CultiVerse workflow was transferable to other art forms, such as Japanese Haiga and Western religious arts. 
E5 suggested extending this workflow to the film and multimedia industries to decode visual and auditory languages in movies, dramas, and plays. 
This could enhance audience engagement by analyzing how symbolic elements, like bamboo or the money tree, reflect different character traits in visual narratives.
Meanwhile, E6 saw potential in applying the workflow to poetry, exploring how a poem in one culture could be analyzed through the lens of other cultures' paintings and open new interpretative dimensions.}

\textbf{Offering users new perspectives.}
\rc{In our user studies, CultiVerse facilitated a richer dialogue with LLMs by eliciting more diverse and insightful user questions than the baseline. 
The difference between CultiVerse and the baseline, which use the same LLMs, seems to stem from the distinct materials used for source culture extraction.
Typically, experts elaborate on artistic elements such as brush techniques and color palettes, but such details may not resonate with all users in a cross-cultural context.
In contrast, CultiVerse's focus on cultural elements encourages users to view art through the lens of cultural norms, which helps them connect more meaningfully with diverse cultural values. 
This approach has increased felt involvement and serendipitous discoveries by exploring various symbolic elements.
We recommend that future systems should prioritize concept-driven exploration over fixed narratives to enhance user engagement across broader audiences.}

\textbf{Providing different interpretation spaces.}
\rc{CultiVerse introduces cultural norms as a means to art appreciation.
This approach enhances user's level of comprehension in TCPs by encouraging them to consider not just ``what'' is depicted but also ``why'' and ``how'' it is presented.
Participants who were well-versed in the source culture reported gaining more insights, suggesting that familiarity with the subject enhances LLM interaction.
However, those less acquainted or from different cultural backgrounds do not necessarily gain ``fewer'' insights but also encounter unique serendipitous findings.
For instance, P6 created Nepali poems inspired by traditional Chinese paintings, highlighting a unique cultural exchange. 
Although introducing cultural norms makes the system's learning curve steeper, this complexity is crucial for facilitating the sophisticated use of LLMs. 
To assist users, we suggest starting with example-driven explorations to ease them into the system.}

\textbf{Enhancing user accessibility.}
\rc{We noticed that LLMs struggle with cultural awareness (\eg stereotypes) and bridging language and cultural barriers for less dominant cultures.
We have leveraged the innate cultural knowledge of users to reduce these challenges, enabling them to connect with unfamiliar cultures using their prior knowledge. 
CultiVerse's streamlined workflow aids users in cross-cultural comparisons and achieving satisfactory outcomes. 
It offers a one-stop, easy-to-use interface that supports context-based exploration with reference-based prompt engineering~\cite{feng2024promptmagician}.
These preset prompts help maintain focus and serve as conversation starters, easing users into interactions with the system, especially those new to LLMs.
Moreover, the accurate information from the TCP-CND dataset ensures reliable results even when LLMs underperform. 
Incorporating multimodal outputs also enhances the novelty and aesthetics of the system.}

\subsection{Limitations and future work}
\rc{In our study of cross-cultural understanding through art, several limitations emerged that need addressing.
Firstly, our research did not consider participants who have bi-cultural or bilingual backgrounds, whose perspectives could enrich our findings. 
Additionally, the diversity and quantity of participants were limited, and we assumed that participants had a strong understanding of their own culture, which may not always be the case. 
Going forward, we plan to use crowdsourcing platforms for broader recruitment in future studies.}

\rc{E3, E6, and E7 suggested that the current system struggles to capture the deeper symbolic meanings essential for comprehensive cultural understanding. 
Accurately identifying subtle symbols remains challenging. For instance, a hooded falcon represents the opposite meaning of a flying one, but identifying a hooded falcon is extremely difficult. 
This requires more sophisticated model training in recognizing such elements. 
Moreover, improving the system's ability to process additional contextual information like historical, geographical, and textual details on artworks is promising for a richer semantic and multimodal interpretation.
Our approach to validating cultural norms using sentiment-based emotions needs improvement, especially in handling composite elements.
E6 also emphasized the system's shortcomings in aligning historical and cultural contexts, with LLMs often misinterpreting historical contexts through contemporary cultural perspectives.
This mismatch and the concerns noted by P2 about stereotypes are crucial areas for future development.}


	

\bibliographystyle{src/abbrv-doi-hyperref}




\section*{Figure Credits}
\label{sec:figure_credits}

\cref{fig:teaser}B3, \cref{fig:teaser}B5, \cref{fig:case2_2}D, \cref{fig:case2_2}E1, and \cref{fig:case2_2}E2 are generated by OpenAI's GPT4-Turbo.



\bibliography{template}


\appendix 

\end{document}